\documentclass{appolb}
\usepackage{amsfonts}

\begin{document}

\eqsec 
\pagestyle{plain}
\title{Distributed-Order Fractional Kinetics
\thanks{Presented at the 16th Marian Smoluchowski Symposium 
on Statistical Physics: Fundamentals and Applications,
September 6-11, 2003}}
\author{
I.M. Sokolov
\address{Institut f\"{u}r Physik, Humboldt-Universit\"{a}t zu Berlin, \\
Newtonstr. 15, D-12489 Berlin, Germany}
\and A.V. Chechkin
\address{Institute for Theoretical Physics, National Science Center\\
"Kharkov Institute of Physics and Technology"\\
Akademicheskaya st.1, 61108 Kharkov, Ukraine}
\and J. Klafter
\address{School of Chemistry, Sackler Faculty of Exact Sciences,\\
Tel Aviv University, Tel Aviv 69978, Israel}}
\maketitle

\begin{abstract}
Fractional diffusion equations are widely used to describe anomalous
diffusion processes where the characteristic displacement scales as a power
of time. For processes lacking such scaling the corresponding
description may be given by distributed-order equations. In the present
paper we consider different forms of distributed-order fractional kinetic
equations and investigate the effects described by different classes of such
equations. In particular, the equations describing accelerating and
decelerating subdiffusion, as well as the those describing accelerating
and decelerating superdiffusion are presented. 
\end{abstract} 


\section{Introduction}

The diffusion equation proposed by Adolf Fick almost 150 years ago is a
partial differential equation of parabolic type, with the first temporal
derivative on its l.h.s. and the second spatial derivative on its r.h.s..
The corresponding equation can be put down both for the particles'
concentration in a diffusing cloud and for a probability of a single
particle's position: 
\begin{equation}
\frac{\partial }{\partial t}p(x,t)=K\frac{\partial ^{2}}{\partial x^{2}}%
p(x,t)
\end{equation}

The mean squared particle's displacement from its initial position given by
the solution of this equation grows linearly in time: $\left\langle
x^{2}(t)\right\rangle =2Kt$. This scaling behavior follows immediately
from the structure of Fick's equation, being second order in spatial
coordinate and first order in time: Changing the spatial scale by a factor
of 2 corresponds to changing the time scale by a factor of 4.

In complex systems this kind of behavior is often violated, being replaced by an
anomalous diffusion relationship, 
\begin{equation}
\left\langle x^{2}(t)\right\rangle \propto t^{\beta },  \label{1.2}
\end{equation}
with $\beta \neq 1$, or (in absence of the second moment) by related
forms where the lower, not necessarily integer, moment of the distribution
scales as a function of time. The continuous description of anomalous
diffusion is given by generalizations of Fick's scheme based on fractional
derivatives \cite{SKB02}. The fractional generalizations of the diffusion equation
may have either a corresponding fractional derivative instead of the
whole-number one, or an additional fractional derivative on the ''wrong''
side of the equation. In what follows we refer to these two possibilities
as to "normal" and "modified" forms of a fractional diffusion equation. 
An example of the "normal" form is an equation for
superdiffusion with the Riesz fractional spatial derivative instead of the
second derivative on the r.h.s. \cite{MK00}. The situation with an additional derivative
on the ''wrong'' side is exemplified by the standard fractional diffusion
equation for subdiffusion \cite{MBK99}. We note that the two forms, i.e. the form with
a Caputo derivative on the l.h.s. for subdiffusive processes, and a form
with an additional spatial derivative on the l.h.s. for a superdiffusive
process, are equivalent to the commonly used ones. This equivalence will be
discussed in detail in the Section 2 of the present work, introducing the
corresponding fractional operators.

Many physical processes, however, lack power-law scaling, Eq.(\ref{1.2}), over
the whole time-domain. These processes can not be characterized by a single
scaling exponent $\beta $. Examples of processes
lacking scaling include several cases of decelerating subdiffusion
(e.g. the Sinai diffusion \cite{Sin82}) and decelerating superdiffusion (as
exemplified by truncated Levy-flights \cite{ManSte}). Such processes can be 
described by derivatives of \textit{distributed} order \cite{SCK,SCK04}, introduced by
Caputo \cite{Cap01}. 
The equivalence between the different forms of fractional diffusion equations 
is lost in this case: Different forms of the distributed-order fractional equations describe
different situations. Thus, the equations with the distributed-order
derivatives on the "proper" side describe processes getting more anomalous in
course of the time (accelerating superdiffusion and decelerating
subdiffusion) \cite{CGS02}, while the equations with the additional distributed-order on
the "wrong" side describe the situations getting less anomalous
(decelerating superdiffusion and accelerating subdiffusion). As an example a special 
model with two fractional derivatives of different orders is used throughout the work.

In Section 2 and 3 we discuss the four forms of fractional diffusion
equations for the processes showing scaling behavior. These forms correspond
to possible permutations of fractional temporal/spatial derivative on the
"proper"/"wrong" side of the equation. Three of the four forms discussed
below are well-known. The fourth one, to our best knowledge, was not
previously discussed. We first turn to the forms pertinent to temporal
fractional equations, i.e. to subdiffusion, and then to the
superdiffusive case. Sections 4 and 5 are devoted to the distributed-order
generalizations of the corresponding equations. The results are 
summarized in Section 6.

\section{Two forms of time fractional diffusion equations}

\subsection{Riemann-Liouville form}

The time fractional diffusion equation (TFDE) in the Riemann-Liouville form
(RL - form), which in our terminology corresponds to a "modified" form of the 
fractional diffusion equation, reads \cite{MK00}: 
\begin{equation}
\frac{\partial }{\partial t}p(x,t)=K_{\beta}\:_0D_{t}^{1-\beta }\frac{\partial
^{2}}{\partial x^{2}}p(x,t),  \label{2.1}
\end{equation}
$p(x,t=0)=\delta (x)$, $0<\beta\leq 1$.

Here $K_{\beta }$ is a positive constant, $[K_{\beta }]=\mathrm{cm}^{2}/%
\mathrm{sec}^{\beta }$, and $_{0}D_{t}^{\mu }$ is the Riemann-Liouville
fractional derivative on the right semi-axis, which, for a ''sufficiently
well-behaved'' function $\phi (t)$ is defined as follows:

\begin{equation}
_{0}D_{t}^{\mu }\phi =\frac{d}{dt}J^{1-\mu }\phi =\frac{1}{\Gamma (1-\mu )}%
\frac{d}{dt}\int_{0}^{t}d\tau \frac{\phi (\tau )}{(t-\tau )^{\mu }},\qquad
0\leq \mu <1,  \label{2.2}
\end{equation}
where $J^{\alpha }\phi (t)=\frac{1}{\Gamma (\alpha )}\int_{0}^{t}d\tau
(t-\tau )^{\alpha -1}\phi (t)$, $t>0$, $\alpha \in \mathbb{R}^{+}$ is the
Riemann-Liouville fractional integral of the order $\alpha $. In what
follows, we omit the subscript ''0'', for brevity.

Applying the Laplace transform, 
\[
\tilde{\phi}(s)\equiv L\left\{ \phi (t)\right\} =\int_{0}^{\infty
}dte^{-st}\phi (t), 
\]
and Fourier-transform, 
\[
g(k)\equiv \Phi \left\{ g(x)\right\} =\int_{-\infty }^{\infty
}dxe^{ikx}g(x) 
\]
in succession,
and using the Laplace transform of the Riemann-Liouville derivative (\ref
{2.2}), 
\begin{equation}
L\left\{ D_{t}^{\mu }\phi (t)\right\} =s^{\mu }\tilde{\phi}(s)  \label{2.4}
\end{equation}
we get from Eq.(\ref{2.1}) the form of the Laplace-transformed
characteristic function $f(k,t)$ of the distribution $p(x,t)$: 
\begin{equation}
\tilde{f}(k,s)=\frac{s^{\beta -1}}{s^{\beta }+K_{\beta }k^{2}}.  \label{2.5}
\end{equation}

\subsection{Caputo form}

The TFDE in the Caputo form (C - form, corresponding to a "normal" form of the
fractional diffusion equation) is written as follows: 
\begin{equation}
\frac{\partial ^{\beta }}{\partial t^{\beta }}p(x,t)=K_{\beta }\frac{%
\partial ^{2}}{\partial x^{2}}p(x,t),  \label{2.6}
\end{equation}
$p(x,0)=\delta (x)$, where $K_{\beta }$ is the same constant as in Eq.(\ref
{2.1}), and the time fractional derivative of order $\beta $, $0<\beta <1$
is understood in the Caputo sense \cite{GM97},

\begin{equation}
\frac{\partial ^{\beta }}{\partial t^{\beta }}\phi \equiv D_{*t}^{\beta
}\phi (t)=J^{1-\beta }\frac{d}{dt}\phi =\frac{1}{\Gamma (1-\beta )}%
\int_{0}^{t}d\tau (t-\tau )^{-\beta }\frac{d}{d\tau }\phi (\tau ).
\label{2.7}
\end{equation}
Here, the sequence of temporal integration and differentiation 
is reversed with respect to a Riemann-Liouville operator.
Recalling the Laplace transform of the Caputo derivative, 
\begin{equation}
L\left\{ \frac{d^{\mu }\phi }{dt^{\mu }}\right\} =s^{\mu }\tilde{\phi}%
(s)-s^{\mu -1}\phi (0),  \label{2.8}
\end{equation}
$0<\mu <1$, and making the Fourier-Laplace transform of Eq.(\ref{2.6}) we
again arrive at Eq.(\ref{2.5}). Thus, both forms of TFDE are equivalent.

In the literature, the third form of TFDE is mentioned, this one also
using the Riemann - Liouville derivative \cite{SZ97}: 
\begin{equation}
D_{t}^{\beta }p(x,t)=K_{\beta }\frac{\partial ^{2}}{\partial x^{2}}p(x,t)+%
\frac{t^{-\beta }}{\Gamma (1-\beta )}p(x,0),  \label{2.9}
\end{equation}
$0<\beta <1$.  The equivalence of the RL- and this third forms can be easily shown
by applying $\int_{0}^{t}dt^{\prime }...$ to both sides of Eq.(\ref{2.1})
and using the Riemann - Liouville derivative of a constant, $D_{t}^{\beta
}1=t^{-\beta }/\Gamma (1-\beta )$.
The equivalence of the "normal" (C-) and the third forms can also be shown easily,
if one uses the relation between the Riemann-Liouville and Caputo
derivatives which can be obtained straightforwardly: 
\begin{equation}
D_{t}^{\beta }\phi (t)=\frac{\partial ^{\beta }}{\partial x^{\beta }}\phi +%
\frac{t^{-\beta }}{\Gamma (1-\beta )}\phi (0)  \label{2.10}
\end{equation}
with $0<\beta <1$. Then, starting, e.g., from the C - form of TFDE, Eq.(\ref
{2.6}), and using Eq.(\ref{2.10}) we immediately arrive at Eq.(\ref{2.9}).

In this paper we are interested in the mean squared displacement (MSD) given by
\begin{equation}
\left\langle x^{2}(t)\right\rangle =L^{-1}\left\{ \left( -\frac{\partial ^{2}%
\tilde{f}}{\partial k^{2}}\right) _{k=0}\right\}.  \label{2.11}
\end{equation}
From Eq.(\ref{2.5}) we get: 
\begin{equation}
\left\langle x^{2}(t)\right\rangle =L^{-1}\left\{ 2K_{\beta }s^{-\beta
-1}\right\} =\frac{2K_{\beta }t^{\beta }}{\Gamma (1+\beta )}.  \label{2.12}
\end{equation}

\section{ Two forms of space fractional diffusion equations}

\subsection{The "normal" form}

The "normal" form of space fractional diffusion equation reads as:
\begin{equation}
\frac{\partial p}{\partial t}=K_{\alpha }\frac{\partial ^{\alpha }p}{%
\partial \left| x\right| ^{\alpha }},  \label{3.1}
\end{equation}
$p(x,t=0)=\delta (x)$, $0<\alpha \leq 2$, where $K_{\alpha }$ is a positive
constant, $[K_{\alpha }]=\mathrm{cm}^{\alpha }/\mathrm{sec}$, and the Riesz
fractional derivative $\partial ^{\alpha }/\partial \left| x\right| ^{\alpha
}$ (we adopt here the notation introduced in \cite{SZ97}) is defined for a
"sufficiently well-behaved" function $f(x)$
through the Liouville - Weil derivatives \cite{SKM93}:
\begin{equation}
\frac{d^{\alpha }}{d\left| x\right| ^{\alpha }}f(x)=\left\{ 
\begin{array}{ll}
-\frac{1}{2\cos (\pi \alpha /2)}\left[ D_{+}^{\alpha }+D_{-}^{\alpha }\right]
& \alpha \neq 1 \\ 
-\frac{d}{dx}\hat{H}f(x) & \alpha =1
\end{array}
\right. ,  \label{3.2}
\end{equation}
where $D_{\pm }^{\alpha }$ are the left - and right side Liouville - Weil
derivatives, 
\begin{eqnarray}
D_{+}^{\alpha }\phi &=&\frac{1}{\Gamma (2-\alpha )}\frac{d^{2}}{dx^{2}}%
\int_{-\infty }^{x}\frac{\phi (\xi )d\xi }{(\xi -x)^{\alpha -1}},  \nonumber
\\
D_{-}^{\alpha }\phi &=&\frac{1}{\Gamma (2-\alpha )}\frac{d^{2}}{dx^{2}}%
\int_{x}^{\infty }\frac{\phi (\xi )d\xi }{(x-\xi )^{\alpha -1}}
\end{eqnarray}
for $0<\alpha <2$, $\alpha \neq 1$ (for $\alpha=1$ $D_{\pm }^{1}=\pm d/dx$), and $H$
is the Hilbert transform operator, 
\[
H\phi =\frac{1}{\pi }\int_{-\infty }^{\infty }\frac{\phi (\xi )d\xi }{x-\xi }%
. 
\]
In Fourier space the operators of fractional derivatives have a simple form: 
\begin{equation}
\Phi (D_{\pm }^{\alpha }\phi )=\int_{-\infty }^{\infty }dx\exp (ikx)D_{\pm
}^{\alpha }\phi =(\mp ik)^{\alpha } \phi(k),  \label{3.4}
\end{equation}
where
\[
(\mp ik)^{\alpha }=\left| k\right| ^{\alpha }\exp \left( \mp \frac{\alpha
\pi i}{2}\mathrm{sign}\,k\right). 
\]
Since 
\begin{equation}
\Phi (H\phi )=i\mathrm{sign}\,k \phi(k)  \label{3.5}
\end{equation}
then, with the use of Eqs.(\ref{3.2}) - (\ref{3.5}) we get the expression,
which is valid for the Fourier transform of the Riesz fractional derivative
for all values of $\alpha$: 
\begin{equation}
\Phi \left( \frac{d^{\alpha }\phi }{d\left| x\right| ^{\alpha }}\right)
=-\left| k\right| ^{\alpha }\phi(k).  \label{3.6}
\end{equation}
Applying the Fourier-transform to Eq.(\ref{3.1}), and noting Eq.(\ref{3.6}),
we get the characteristic function for the PDF of L\'{e}vy flights, 
\begin{equation}
f(k,t)=\exp \left( -D\left| k\right| ^{\alpha }t\right).  \label{3.7}
\end{equation}

\subsection{The "modified" form}

Let us turn to the fractional equation for superdiffusion with the
additional spatial derivative on its l.h.s. 
\begin{equation}
\frac{\partial ^{2-\alpha }}{\partial \left| x\right| ^{2-\alpha }}\frac{%
\partial p}{\partial t}=-K_{\alpha }\frac{\partial ^{2}}{\partial x^{2}}p,
\label{3.8}
\end{equation}
where $K_{\alpha }$ is the same as in Eq.(\ref{3.1}). Note the minus sign in
Eq.(\ref{3.8}).
This sign gets clear when turning to a Fourier representation: applying
Fourier transform and using Eq.(\ref{3.6}), we arrive at the characteristic
function (\ref{3.7}). Thus, both forms of the space fractional diffusion
equation, Eqs.(\ref{3.1}) and (\ref{3.8}), are equivalent.

Since the mean square displacement diverges for L\'{e}vy flights, their
anomalous nature can be characterized by a typical displacement $%
\delta x$ of the diffusing particle, 
\begin{equation}
\delta x\propto \left\langle \left| x\right| ^{q}\right\rangle ^{1/q},\qquad
q<\alpha ,  \label{3.9}
\end{equation}
which differs, of course, from the MSD discussed in Eq.(\ref{2.11}).
To get the $q$-th moment we use the following expression \cite{Zol86}: 
\begin{equation}
\left\langle \left| x\right| ^{q}\right\rangle =\frac{2}{\pi }\Gamma
(1+q)\sin \frac{\pi q}{2}\int_{0}^{\infty }dk\left( 1-\mbox{Re}f(k,t)\right)
k^{-q-1}.  \label{3.10}
\end{equation}
Inserting Eq.(\ref{3.7}) into Eq.(\ref{3.10}) and changing to a new variable, 
$\xi =K_{\alpha }k^{\alpha }t$, we obtain
\begin{equation}
\int_{0}^{\infty }dk...=\frac{(K_{\alpha }t)^{1/\alpha }}{\alpha }%
\int_{0}^{\infty }d\xi (1-e^{-\xi })\xi ^{-q/\alpha -1}=\frac{(K_{\alpha
}t)^{q/\alpha }}{q}\Gamma \left( 1-\frac{q}{\alpha }\right),  \label{3.11}
\end{equation}
and, thus 
\begin{equation}
\left\langle \left| x\right| ^{q}\right\rangle =C(q,\alpha )(K_{\alpha
}t)^{q/\alpha },\qquad q<\alpha.  \label{3.12}
\end{equation}
Here 
\begin{equation}
C(q,\alpha )=\frac{2}{\pi q}\sin \left( \frac{\pi q}{2}\right) \Gamma
(1+q)\Gamma \left( 1-\frac{q}{\alpha }\right) .  \label{3.13}
\end{equation}
Note that for $a=q=2$ Eqs.(\ref{3.12}), (\ref{3.13}) give 
\[
\left\langle x^{2}\right\rangle =2Kt. 
\]

\section{Distributed-order time fractional diffusion equations}

\subsection{Distributed-order time fractional diffusion equation in the RL
form}

The fractional diffusion equation with a distributed-order
Riemann-Liouville derivative reads: 
\begin{equation}
\frac{\partial p}{\partial t}=\int_{0}^{1}d\beta w(\beta )K(\beta
)D_{i}^{1-\beta }\frac{\partial ^{2}p}{\partial x^{2}},  \label{4.1}
\end{equation}
$p(x,0)= \delta(x)$, where $K(\beta )=K\tau ^{1-\beta }$, 
$[K]=\mathrm{cm}^{2}/\mathrm{sec}$, $%
[\tau ]=\mathrm{sec}$, $w(\beta )$ is a dimensionless non-negative function,
which should fulfill $\int_{0}^{1}d\beta w(\beta )=1$. If we set $w(\beta
)=\delta (\beta -\beta _{0})$, $0<\beta _{0}<1$, then we arrive at time
fractional diffusion equation in the RL form, see Eq.(\ref{2.1}), where $%
K_{\beta }=K\tau ^{1-\beta _{0}}$.

We now prove that the solution of Eq.(\ref{4.1}) is a PDF. The derivation
here follows the method used in \cite{Sok01}. Its aim is to show that
the random process whose PDF obeys Eq.(\ref{4.1}) is subordinated to the
Wiener process. Making a Fourier-Laplace transform of Eq.(\ref{4.1}) and
using Eq.(\ref{2.4}) we get 
\begin{equation}
\tilde{f}(k,s)=\frac{1}{sI_{RL}\left( I_{RL}^{-1}+k^{2}K\tau \right) },
\label{4.2}
\end{equation}
where 
\begin{equation}
I_{RL}(s\tau )=\int_{0}^{1}d\beta (s\tau )^{-\beta }w(\beta ).  \label{4.3}
\end{equation}
We rewrite Eq.(\ref{4.2}) as follows: 
\begin{equation}
\tilde{f}(k,s)=\frac{1}{sI_{RL}}\int_{0}^{\infty }du\exp \left[ -u\left(
I_{RL}^{-1}+k^{2}K\tau \right) \right] =\int_{0}^{\infty }due^{-uk^{2}K\tau }%
\tilde{G}_{RL}(u,s)  \label{4.4}
\end{equation}
where 
\begin{equation}
\tilde{G}_{RL}(u,s)=\frac{1}{sI_{RL}(s\tau )}\exp \left[ -\frac{u}{%
I_{RL}(s\tau )}\right]  \label{4.5}
\end{equation}
is a Laplace transform of a function whose properties will be specified
below. Now, with the help of Eqs.(\ref{4.4}) and (\ref{4.5}) the PDF $p(x,t)$
can be written as 
\begin{eqnarray}
p(x,t) &=&\int_{-\infty }^{\infty }\frac{dk}{2\pi }e^{-ikx}\int_{Br}\frac{ds%
}{2\pi i}e^{st}\int_{0}^{\infty }due^{-uk^{2}K\tau }\tilde{G}_{RL}(u,s)= 
\nonumber \\
&=&\int_{0}^{\infty }du\frac{e^{-x^{2}/4uK\tau }}{\sqrt{4\pi uK\tau }}%
G_{RL}(u,t)  \label{4.6}
\end{eqnarray}
where $Br$ denotes the Bromwich integration contour.

In order to prove the positivity of $p(x,t)$ we demonstrate that the
function $G_{RL}(u,t)$ is the PDF providing the subordination
transformation from time scale $t$ to time scale $u$, that is, $G_{RL}(u,t)$
is positive and normalized with respect to $u$ for any $t$. At first we
demonstrate normalization, 
\begin{equation}
\int_{0}^{\infty }duG_{RL}(u,t)=L^{-1}\left\{ \int_{0}^{\infty }du\frac{%
e^{-u/I_{RL}}}{sI_{RL}}\right\} =L^{-1}\left\{ \frac{1}{s}\right\} =1.
\label{4.7}
\end{equation}
Now, to prove the positivity of function $G_{RL}(u,t)$, according to the
Bernstein theorem \cite{Fel71}, it is enough to show that  $\tilde{G}_{RL}(u,s)$, 
Eq.(\ref{4.5}), is
completely monotonic as a function of $s$ on positive real axis, i.e., it is
positive and the signs of its derivatives alternate. We do this for the
special case when 
\begin{equation}
w(\beta )=B_{1}\delta (\beta -\beta _{1})+B_{2}\delta (\beta -\beta _{2})
\label{4.8}
\end{equation}
with $0<\beta _{1}<\beta _{2}\leq 1$, $B_{1}>0$, $B_{2}>0$, $B_{1}+B_{2}=1$.
This choice allows us to show in a simple way the property of the diffusive
behavior governed by distributed-order diffusion equations. This is why this
case is repeatedly discussed in our article. 

Inserting Eq.(\ref{4.8})
into Eq.(\ref{4.3}) we have
\begin{equation}
I_{RL}(s\tau )=b_{1}s^{-\beta _{1}}+b_{2}s^{-\beta _{2}}  \label{4.9}
\end{equation}
where $b_{1}=B_{1}/\tau ^{\beta _{1}}$, $b_{2}=B_{2}/\tau ^{\beta _{2}}$.
From Eq.(\ref{4.5}) it follows that $\tilde{G}_{RL}(u,s)$ is a product
of two functions: $\tilde{G}_{RL}(u,s)=\phi _{1}(s)\phi _{2}(u,s)$ with
\begin{equation}
\phi _{1}=\frac{1}{sI_{RL}(s\tau )},\qquad \phi _{2}=\exp \left[ -\frac{u}{%
I_{RL}(s\tau )}\right].
\label{4.10}
\end{equation}
We will prove that both functions $\phi _{1}$ and $\phi _{2}$ are completely
monotonic, and therefore, as a product, $\tilde{G}_{RL}(u,s)$ is completely
monotonic too.

Let us start from $\phi _{1}$. It can be rewritten as $\phi _{1}=\phi (h(s))$%
, where $\phi (y)=1/y$\ is a completely monotonic function, and $%
h(s)=b_{1}s^{1-\beta _{1}}+b_{2}s^{1-\beta _{2}}$ is a positive function
with a completely monotonic derivative (this is evident by direct
inspection). Therefore, $\phi _{1}(s)$ is completely monotonic according to Criterion 2 in
 \cite{Fel71}, Chapter XIII, \S4.

Now turn to $\phi _{2}$. Again, it can be rewritten as $\phi _{2}=\phi (\psi
(s))$, where $\phi (y)=\exp (-uy)$ is a completely monotonic function.
According to the same Criterion, it is enough to show that the positive
function 
\begin{equation}
\psi (s)=\frac{1}{I_{RL}(s\tau )}=\frac{1}{b_{1}s^{-\beta
_{1}}+b_{2}s^{-\beta _{2}}}=\frac{s^{\beta _{2}}}{b_{2}}\frac{1}{1+\frac{%
b_{1}}{b_{2}}s^{\beta _{2}-\beta _{1}}}  \label{4.11}
\end{equation}
possesses completely monotonic derivative. By differentiating Eq.(\ref{4.11}%
) we get 
\begin{eqnarray}
\psi ^{\prime }(s) &=&\frac{s^{\beta _{2}-1}}{b_{2}}\frac{1}{1+\frac{b_{1}}{%
b_{2}}s^{\beta _{2}-\beta _{1}}}-\frac{s^{\beta _{2}}}{b_{2}}\frac{\frac{%
b_{1}}{b_{2}}(\beta _{2}-\beta _{1})s^{\beta _{2}-\beta _{1}-1}}{\left( 1+%
\frac{b_{1}}{b_{2}}s^{\beta _{2}-\beta _{1}}\right) ^{2}}=  \nonumber \\
&=&\frac{\beta _{2}s^{\beta _{2}-1}}{b_{2}}\frac{1}{1+\frac{b_{1}}{b_{2}}%
s^{\beta _{2}-\beta _{1}}}\frac{1+\frac{b_{1}\beta _{1}}{b_{2}\beta _{2}}%
s^{\beta _{2}-\beta _{1}}}{1+\frac{b_{1}}{b_{2}}s^{\beta _{2}-\beta _{1}}}.
\label{4.12}
\end{eqnarray}
Denoting $\xi =\beta _{2}-\beta _{1}$, $0<\xi <1$, we note that this
function is a product of three completely monotonic functions, and thus is
itself completely monotone, too. Indeed,
(i) the first function is $s^{\beta _{2}-1}$being a negative power of $s$;
(ii) the second one is completely monotonic since it has the form $g(h(s))$,
with $g(y)=1/(1+y)$ being completely monotone, and $h(\xi )=s^{\xi }$ is a
positive function with a completely monotone derivative;
(iii) the last function has the same form with $g(y)=(1+cy)/(1+dy)$, where $%
0<c<d$ with $c=b_{1}\beta _{1}/b_{2}\beta _{2}$, $d=b_{1}/b_{2}$. The
function $g(y)$ is completely monotonic since its $n$-th derivative
(obtained using the Leibnitz rule) has the form 
\[
g^{(n)}(y)=\frac{(-1)^{n-1}(c-d)n!d^{n-1}}{(1+dy)^{n+1}}, 
\]
and the signs of the successive derivatives alternate. Thus, we have proved
that $\psi ^{\prime }(s)$, Eq.(\ref{4.12}), is a completely monotonic
function. Therefore the exponential $\phi _{2}$, Eq.(\ref{4.10}), is a
completely monotonic function, too, and the whole function, $\tilde{G}%
_{RL}(u,s)$, is completely monotone as a product of two completely
monotonic functions, $\phi _{1}$ and $\phi _{2}$. Therefore, the function $%
G_{RL}(u,t)$ is a PDF and, according to Eq.(\ref{4.6}), the function $p(x,t)$
is a PDF, too. Thus, we have proved that the solution of the distributed
order time fractional diffusion equation in the Riemann-Liouville form is a
PDF.

We are interested in the MSD which is given by
\begin{equation}
\left\langle x^{2}(t)\right\rangle =L^{-1}\left\{ \left( -\frac{\partial ^{2}%
\tilde{f}}{\partial k^{2}}\right) _{k=0}\right\} =K\tau L^{-1}\left\{ \frac{%
2I_{RL}(s\tau )}{s}\right\} .  \label{4.13}
\end{equation}
For the particular case (\ref{4.8}) we get, by inserting Eq.(\ref{4.9}) into
Eq.(\ref{4.13}) and making an inverse Laplace transform,
\begin{equation}
\left\langle x^{2}(t)\right\rangle =\frac{2\kappa _{1}}{\Gamma (1+\beta _{1})%
}\left( \frac{t}{\tau }\right) ^{\beta _{1}}+\frac{2\kappa _{2}}{\Gamma
(1+\beta _{2})}\left( \frac{t}{\tau }\right) ^{\beta _{2}}  \label{4.14}
\end{equation}
where $\kappa _{1}=B_{1}K\tau $ and $\kappa _{2}=B_{2}K\tau $. Since $%
0<\beta _{1}<\beta _{2}\leq 1$, at small times the first term in the r.h.s.
of Eq.(\ref{4.14}) prevails, whereas at large times the second one
dominates. Thus, the overall behavior corresponds to accelerating subdiffusion.

\subsection{Distributed-order time fractional diffusion equation in the C -
form}

The distributed-order time fractional diffusion equation in the "normal"
form can be written as
\begin{equation}
\int_{0}^{1}d\beta \tau ^{\beta -1}w(\beta )\frac{\partial ^{\beta }p}{%
\partial t^{\beta }}=K\frac{\partial ^{2}p}{\partial x^{2}},  \label{4.15}
\end{equation}
$p(x,0)=\delta (x)$, where $\tau $ is a positive constant representing some
characteristic time of the problem (vide infra), $[\tau ]=\mathrm{sec}$, $K$
is the diffusion coefficient, $[K]=\mathrm{cm}^{2}/\mathrm{sec}$, $w(\beta )$
is a dimensionless non-negative function, and the time fractional derivative
of order $\beta $ , $0<\beta <1$ is understood in the Caputo sense, Eq.(%
\ref{2.7}). Note the difference between Eq.(\ref{4.15}) and Eq.(\ref{4.1}).

If we set $w(\beta )=\delta (\beta -\beta _{0})$, $0<\beta _{0}<1$, we
arrive at the "normal" form of a fractional diffusion equation, Eq.(\ref{2.6}), 
with $\beta=\beta _{0}$ and $K_{\beta }=K\tau ^{1-\beta _{0}}$.

Let us prove that the solution of Eq.(\ref{4.15}) is a PDF. 
Applying the Laplace- and
Fourier-transforms in succession, we get: 
\begin{equation}
\tilde{f}(k,s)=\frac{1}{s}\frac{I_{C}(s\tau )}{I_{C}(s\tau )+k^{2}K\tau },
\label{4.16}
\end{equation}
where 
\begin{equation}
I_{C}(s\tau )=\int_{0}^{1}d\beta (s\tau )^{\beta }w(\beta ).  \label{4.17}
\end{equation}
We rewrite Eq.(\ref{4.16}) in the form analogous to Eq.(\ref{4.4}),
\begin{equation}
\tilde{f}(k,s)=\frac{I_{C}}{s}\int_{0}^{\infty }du\,e^{-u\left[
I_{c}+k^{2}K\tau \right] }=\int_{0}^{\infty }du\,e^{-uk^{2}K\tau }\tilde{G}%
_{C}(u,s)  \label{4.18}
\end{equation}
where 
\begin{equation}
\tilde{G}_{C}(u,s)=\frac{I_{C}(s\tau )}{s}e^{-uI_{C}(s\tau )}  \label{4.19}
\end{equation}
is the Laplace transform of a function $G_{C}(u,t)$ whose properties will be
specified below. Now, $p(x,t)$ can be written in the form analogous to Eq.(%
\ref{4.6}):
\begin{eqnarray}
p(x,t) &=&\int_{-\infty }^{\infty }\frac{dk}{2\pi }e^{-ikx}\int_{Br}\frac{ds%
}{2\pi i}e^{st}\int_{0}^{\infty }due^{-uk^{2}K\tau }\tilde{G}_{C}(u,s)= 
\nonumber \\
&=&\int_{0}^{\infty }du\frac{e^{-x^{2}/4uK\tau }}{\sqrt{4\pi uK\tau }}%
G_{C}(u,t)  \label{4.20}
\end{eqnarray}
Similar to the RL case, the function $G_{C}(u,t)$ is the PDF providing the
subordination transformation, from time scale $t$ to time scale $u$. Indeed,
at first we note that $G_{C}(u,t)$ is normalized with respect to $u$ for any 
$t$.  Using Eq.(\ref{4.19}) we get 
\begin{equation}
\int_{0}^{\infty }duG_{C}(u,t)=L^{-1}\left\{ \int_{0}^{\infty }du\frac{I_{RL}%
}{s}e^{-u/I_{C}}\right\} =L^{-1}\left\{ \frac{1}{s}\right\} =1.  \label{4.21}
\end{equation}

To prove the positivity of $G_{C}(u,t)$ it is enough to show 
that its Laplace transform is completely monotone
on the positive real axis  \cite{Fel71}. The last statement is proved by
noting that it is a product of two completely monotonic functions, $I_{C}/s$
and $\exp (-uI_{C})$. We again demonstrate this for the particular choice of 
$w(\beta )$, see Eq.(\ref{4.8}). For this choice we obtain from Eq.(\ref{4.17})
\begin{equation}
I_{C}(s)=b_{1}s^{\beta _{1}}+b_{2}s^{\beta _{2}},  \label{4.22}
\end{equation}
where $b_{1}=B_{1}\tau ^{\beta _{1}}$, $b_{2}=B_{2}\tau ^{\beta _{2}}$.
Then, $I_{C}/s$ is completely monotone as a sum of the negative powers of $%
s $, and $I_{C}$ itself is a positive function with a completely monotone
derivative. Thus, $\exp (-uI_{C})$ is also completely monotone. Thus,
we have proved that is completely monotone, and $G_{RL}(u,t)$ is a PDF, according
to the Bernstein theorem.

We are interested in the behavior of the MSD, i.e. in second moment of the PDF. 
Using Eq.(\ref{4.16}), we have 
\begin{equation}
\left\langle x^{2}(t)\right\rangle =L^{-1}\left\{ \left( -\frac{\partial ^{2}%
\tilde{f}}{\partial k^{2}}\right) _{k=0}\right\} =2K\tau L^{-1}\left\{ \frac{%
1}{sI_{C}(s\tau )}\right\} .  \label{4.23}
\end{equation}
Inserting Eq.(\ref{4.22}) into Eq.(\ref{4.23}) one obtains:
\begin{equation}
\left\langle x^{2}(t)\right\rangle =2K\tau L^{-1}\left\{ \frac{1}{s\left(
b_{1}s^{\beta _{1}}+b_{2}s^{\beta _{2}}\right) }\right\} =\frac{2K\tau }{%
b_{2}}L^{-1}\left\{ \frac{s^{-\beta _{1}-1}}{\frac{b_{1}}{b_{2}}+s^{\beta
_{2}-\beta _{1}}}\right\} .  \label{4.24}
\end{equation}
Recalling the Laplace transform of the generalized Mittag-Leffler function 
$E_{\mu ,\nu }(z)$, $\mu >0$, $\nu >0$, which can be conveniently written as
\cite{GM97}
\begin{equation}
L\left\{ t^{\nu -1}E_{\mu ,\nu }(-\lambda t^{\mu })\right\} =\frac{s^{\mu
-\nu }}{s^{\mu }+\lambda },\qquad \mbox{Re}s>\left| \lambda \right| ^{1/\mu },
\label{4.25}
\end{equation}
we get from Eq.(\ref{4.24}): 
\begin{equation}
\left\langle x^{2}(t)\right\rangle =\frac{2K\tau }{b_{2}}t^{\beta
_{2}}E_{\beta _{2}-\beta _{1},\beta _{2}+1}\left( -\frac{b_{1}}{b_{2}}%
t^{\beta _{2}-\beta _{1}}\right).  \label{4.26}
\end{equation}

To obtain asymptotics at small $t$, we use an expansion, which is, in fact the
definition of $E_{\mu ,\nu }(z)$, see \cite{Erd55}, Ch.XVIII, Eq.(19): 
\begin{equation}
E_{\mu ,\nu }(z)=\sum_{n=0}^{\infty }\frac{z^{n}}{\Gamma (\mu n+\nu )},
\label{4.27}
\end{equation}
which yields in the main order for the MSD
\begin{equation}
\left\langle x^{2}(t)\right\rangle =\frac{2K\tau }{B_{2}\Gamma (\beta _{2}+1)%
}\left( \frac{t}{\tau }\right) ^{\beta _{2}}\propto t^{\beta _{2}}.
\label{4.28}
\end{equation}

For large $t$ we use the following expansion valid on the real negative
axis, see \cite{Erd55}, Ch.XVIII, Eq.(21): 
\begin{equation}
E_{\mu ,\nu }(z)=-\sum_{n=0}^{N}\frac{z^{-n}}{\Gamma (-\mu n+\nu )}+O\left(
\left| z\right| ^{-1-N}\right) ,\qquad \left| z\right| \rightarrow \infty
\label{4.29}
\end{equation}
which yields 
\begin{equation}
\left\langle x^{2}(t)\right\rangle =\frac{2K\tau }{B_{1}\Gamma (\beta _{1}+1)%
}\left( \frac{t}{\tau }\right) ^{\beta _{1}}\propto t^{\beta _{1}}.
\label{4.30}
\end{equation}
Since $\beta _{1}<\beta _{2}$, we have retarded subdiffusion. It is
worthwhile to note that the distributed-order equation (\ref{4.15}) with the
Caputo derivative always describes retarding, or slowing-down,
sub-diffusive processes. Indeed, it is clearly seen from Eqs.(\ref{4.22}), (%
\ref{4.23}) that, for $w(\beta )$ given by Eq.(\ref{4.8}), for large $s$
(short times) $I(s\tau )$ behaves as $s^{\beta _{2}}$, so that, due to
the Tauberian theorem, the MSD behaves as $t^{\beta _{2}}$, while at long times
(small $s$) $I(s\tau )\propto s^{\beta _{1}}$ and, respectively, $%
\left\langle x^{2}\right\rangle \propto t^{\beta _{1}}$, $\beta _{1}<\beta
_{2}$. Thus, Eq.(\ref{4.15}) cannot describe the accelerating
sub-diffusive process. The same arguments applied to Eq.(%
\ref{4.1}) with the Riemann - Liouville derivative demonstrate that Eq.(\ref
{4.1}) cannot describe retarded sub-diffusive processes.

\section{Distributed-Order Space Fractional Diffusion Equation}

\subsection{The "normal" form}

We write the distributed-order space fractional diffusion equation in the
"normal" form as
\begin{equation}
\frac{\partial p}{\partial t}=\int_{0}^{2}d\alpha K(\alpha )\frac{\partial
^{\alpha }p}{\partial \left| x\right| ^{\alpha }},\qquad p(x,0)=\delta (x)
\label{5.1}
\end{equation}
where $K(\alpha )$ is a (dimensional) function of the order of the
derivative $\alpha $, and the Riesz space fractional derivative $\partial
^{\alpha }/\partial \left| x\right| ^{\alpha }$ is given by its
Fourier transform, see Eq.(\ref{3.6}). Setting $K(\alpha )=K_{\alpha
_{0}}\delta (\alpha -\alpha _{0})$ we arrive at Eq.(\ref{3.1}). In the
general case $K(\alpha )$ can be represented as
\begin{equation}
K(\alpha )=l^{\alpha -2}Kw(\alpha ),  \label{5.2}
\end{equation}
where $l$ and $K$ are dimensional positive constants, $[l]=\mathrm{cm}$, $%
[K]=\mathrm{cm}^{2}/\mathrm{sec}$ and $w$ is a dimensionless non-negative
function of $\alpha $. The equation for the characteristic
function of Eq.(\ref{5.1}) has the solution 
\begin{equation}
f(k,t)=\exp \left\{ -\frac{Kt}{l^{2}}\int_{0}^{2}d\alpha w(\alpha )(\left|
k\right| l)^{\alpha }\right\} .  \label{5.3}
\end{equation}
Note that the normalization condition, 
\begin{equation}
\int_{-\infty }^{\infty }dxp(x,t)=f(k=0,t)=1  \label{5.4}
\end{equation}
is fulfilled.

Let us consider the simple particular case, 
\begin{equation}
w(\alpha )=A_{1}\delta (\alpha -\alpha _{1})+A_{2}\delta (\alpha -\alpha
_{2}),  \label{5.5}
\end{equation}
where $0<\alpha _{1}<\alpha _{2}\leq 2$, $A_{1}>0$, $A_{2}>0$. Inserting Eq.(%
\ref{5.5}) into Eq.(\ref{5.3}) we have
\begin{equation}
f(k,t)=\exp \left\{ -a_{1}\left| k\right| ^{\alpha _{1}}t-a_{2}\left|
k\right| ^{\alpha _{2}}t\right\}  \label{5.6}
\end{equation}
where $a_{1}=A_{1}K/l^{2-\alpha _{1}}$, $a_{2}=A_{2}K/l^{2-\alpha _{2}}$.
The characteristic function (\ref{5.6}) is a product of two characteristic
functions of L\'{e}vy stable PDFs with the L\'{e}vy indexes $\alpha _{1}$%
, $\alpha _{2}$, and the scale parameters $a_{1}^{1/\alpha _{1}}$ and $%
a_{2}^{1/\alpha _{2}}$, respectively. Therefore, the inverse Fourier
transformation of Eq.(\ref{5.6}) gives the PDF which is a convolution of
two stable PDFs, 
\begin{equation}
p(x,t)=a_{1}^{-\frac{1}{\alpha _{1}}}a_{2}^{-\frac{1}{\alpha _{2}}}t^{-\frac{1}{\alpha
_{1}}-\frac{1}{\alpha _{2}}}\int_{-\infty }^{\infty }dx^{\prime }L_{\alpha
_{1},1}\left( \frac{x-x^{\prime }}{(a_{1}t)^{\frac{1}{\alpha _{1}}}}\right)
L_{\alpha _{2},1}\left( \frac{x^{\prime }}{(a_{2}t)^{\frac{1}{\alpha _{2}}}}\right),
\label{5.7}
\end{equation}
where $L_{\alpha _{{}},1}$ is the PDF of a symmetric L\'{e}vy stable law
given by its characteristic function 
\begin{equation}
\hat{L}_{\alpha ,1}(k)=\exp \left( -\left| k\right| ^{\alpha }\right) .
\label{5.8}
\end{equation}
The PDF given by Eq.(\ref{5.7}) is, obviously, positive, as the convolution
of two positive PDFs. 

The PDF will be also positive, if the function $%
A(\alpha )$ is represented as a sum of $N$ delta-functions multiplied by
positive constants, $N$ is a positive integer. Moreover, if $A(\alpha )$ is
a continuous positive function, then discretizing the integral in Eq.(\ref
{5.1}) by a Riemann sum and passing to the limit we can also conclude on the
positivity of the PDF in the general case.

Let us consider the $q$-th moment, Eq.(\ref{3.10}), at small and large $t$.
After inserting Eq.(\ref{5.6}) into Eq.(\ref{3.10}) we get 
\begin{equation}
\left\langle \left| x\right| ^{q}\right\rangle =\frac{2}{\pi }\Gamma
(1+q)\sin \left( \frac{\pi q}{2}\right) \int_{0}^{\infty }dk\left(
1-e^{-t\phi (k)}\right) k^{-q-1},  \label{5.9}
\end{equation}
where 
\begin{equation}
\phi (k)=a_{1}k^{\alpha _{1}}+a_{2}k^{\alpha _{2}}.  \label{5.10}
\end{equation}

In order to get the $q$-th moment at small $t$, we expand $\exp \left(
-a_{1}\left| k\right| ^{\alpha _{1}}t\right) $ in power series with
subsequent integration over $k$. As the result we have the following
expansion valid at $q<\alpha _{1}$: 
\begin{eqnarray}
\left\langle \left| x\right| ^{q}\right\rangle &=&\frac{2}{\pi q}%
(a_{2}t)^{q/\alpha _{2}}\sin \left( \frac{\pi q}{2}\right) \Gamma
(1+q)\Gamma \left( 1-\frac{q}{\alpha _{2}}\right) \times  \label{5.11} \\
&&\left\{ 1+\frac{1}{\Gamma \left( 1-\frac{q}{\alpha _{2}}\right) }%
\sum_{n=1}^{\infty }\frac{(-1)^{n+1}}{\alpha _{2}n!}a_{1}^{n}a_{2}^{-\frac{%
n\alpha _{1}}{\alpha _{2}}}\Gamma \left( \frac{n\alpha _{1}-q}{\alpha _{2}}%
\right) t^{n\left( 1-\frac{\alpha _{1}}{\alpha _{2}}\right) }\right\} 
\nonumber
\end{eqnarray}
for $t\rightarrow 0$. We recall that the radius $R$ of convergence of a
series $\sum_{n=0}^{\infty }c_{n}x^{n}$ is determined from the equation 
\[
1/R=\lim_{n\rightarrow \infty }\left| c_{n}\right| ^{1/n}. 
\]
Then, by using the Stirling formula, one can see that 
in case of Eq.(\ref{5.11}) $\left| c_{n}\right|
^{1/n}\sim n^{\alpha _{1}/\alpha _{2}-1}\rightarrow 0$ for $n\rightarrow
\infty $, and therefore expansion (\ref{5.11}) is valid at all $t$. The leading
term of a series (\ref{5.11}) gives for the characteristic displacement at
small $t$, 
\begin{equation}
\delta x\sim \left\langle \left| x\right| ^{q}\right\rangle ^{1/q}\propto
t^{1/\alpha _{2}}.  \label{5.12}
\end{equation}
In order to get the $q$-th moment at large $t$ we integrate by parts
the right hand side of Eq.(\ref{5.9}): 
\begin{equation}
\left\langle \left| x\right| ^{q}\right\rangle =\frac{2}{\pi }\Gamma
(1+q)\sin \left( \frac{\pi q}{2}\right) \frac{t}{q}J(t),  \label{5.13}
\end{equation}
where 
\begin{equation}
J(t)=\int_{0}^{\infty }dk\zeta (k)e^{-t\phi (k)}  \label{5.14}
\end{equation}
with
\[
\zeta (k)=\frac{\phi ^{\prime }(k)}{k^{q}}=\frac{a_{1}\alpha _{1}k^{\alpha
_{1}}+a_{2}\alpha _{2}k^{\alpha _{2}}}{k^{q+1}}. 
\]
Since at small $k$ one has $\phi (k)\approx a_{1}k^{\alpha _{1}}$, $\zeta (k)\approx
a_{1}\alpha _{1}k^{\alpha _{1}-q-1}$, then at large $t$ we have exactly the
case of Laplace asymptotic integral, see \cite{Olv74}, Chapter III,
Theorem 7.1. For $J(t)$ one immediately obtains $J(t)=a_{1}^{q/\alpha
_{1}}t^{q/\alpha _{1}-1}\Gamma (1-q/\alpha _{1})$ and 
\begin{equation}
\left\langle \left| x\right| ^{q}\right\rangle \sim \frac{2}{\pi q}%
(a_{1}t)^{q/\alpha _{1}}\sin \left( \frac{\pi q}{2}\right) \Gamma
(1+q)\Gamma \left( 1-\frac{q}{\alpha _{1}}\right) ,\qquad q<\alpha _{1}.
\label{5.15}
\end{equation}
For the characteristic displacement at large $t$ we get 
\begin{equation}
\delta x\sim \left\langle \left| x\right| ^{q}\right\rangle ^{1/q}\propto
t^{1/\alpha _{1}}.  \label{5.16}
\end{equation}
Therefore, at small times the characteristic displacement grows as $%
t^{1/\alpha _{2}}$, whereas at large times it grows as $t^{1/\alpha _{1}}$.
Since $\alpha _{1}<\alpha _{2}$, one encounters an accelerated superdiffusion.

\subsection{The "modified" form}

Let us now consider the conjugated form of the equation, namely one with an
additional distributed-order fractional operator in the l.h.s.: 
\begin{equation}
\int_{0}^{2}d\alpha w(\alpha )l^{2-\alpha }\frac{\partial ^{2-\alpha }}{%
\partial \left| x\right| ^{2-\alpha }}\frac{\partial p}{\partial t}=-K\frac{%
\partial ^{2}p}{dx^{2}},\qquad p(x,0)=\delta (x),  \label{5.17}
\end{equation}
where $l$, $K$ and $w$ have the same meaning as in Eq.(\ref{5.2}). Setting 
$w(\alpha )=\delta (\alpha -\alpha _{0})$ we arrive at Eq.(\ref{3.8})
where $K_{\alpha }\equiv K_{\alpha _{0}}=Kl^{\alpha _{0}-2}$. The equation
for the characteristic function of the solution of Eq.(\ref{5.17}) reads:  
\begin{equation}
f(k,t)=\exp \left[ -\frac{Kt/l^{2}}{\int_{0}^{2}d\alpha w(\alpha )(\left|
k\right| l)^{\alpha }}\right]  \label{5.18}
\end{equation}
(compare with Eq.(\ref{5.3})). 
Note that the normalization condition, Eq.(\ref{5.4}), is fulfilled.

As we did it throughout the present article, let us consider a particular
case of 
\begin{equation}
w(\alpha )=A_{1}\delta (\alpha -\alpha _{1})+A_{2}\delta (\alpha -\alpha
_{2}),  \label{5.19}
\end{equation}
$0<\alpha _{1}<\alpha _{2}\leq 2$. From Eq.(\ref{5.18}) we get for the
characteristic function, 
\begin{equation}
f(k,t)=\exp \left[ -\frac{t}{\frac{a_{1}}{\left| k\right| ^{\alpha _{1}}}+%
\frac{a_{2}}{\left| k\right| ^{\alpha _{2}}}}\right] ,  \label{5.20}
\end{equation}
where $a_{1,2}=A_{1,2}l^{2-\alpha _{1,2}}/K$. The proof of
non-negativity of the PDF given by inverse Fourier transform of Eq.(\ref
{5.20}) follows along the same lines as for the accelerating subdiffusion
case.

The $q$-th moment is given by Eq.(\ref{5.9}), where 
\begin{equation}
\phi (k)=\frac{1}{a_{1}k^{-\alpha _{1}}+a_{2}k^{-\alpha _{2}}}.  \label{5.21}
\end{equation}
We insert Eq.(\ref{5.21}) into Eq.(\ref{5.9}) and pass to a new variable $%
\xi =tk^{\alpha _{1}}/\alpha _{1}$. For the integral over $k$ we get 
\begin{eqnarray}
\int_{0}^{\infty }dk... &=&\left( \frac{t}{a_{1}}\right) ^{q/\alpha _{1}}%
\frac{1}{\alpha _{1}}\times  \label{5.22} \\
&&\int_{0}^{\infty }d\xi \xi ^{-1-q/\alpha _{1}}\left\{ 1-\exp \left[ -\frac{%
\xi }{1+\frac{a_{2}}{a_{1}}\xi ^{1-\frac{\alpha _{2}}{\alpha _{1}}}\left( 
\frac{t}{a}\right) ^{\frac{\alpha _{2}}{\alpha _{1}}-1}}\right] \right\}. 
\nonumber
\end{eqnarray}
At small $t$ we can neglect the term with $t$ in the denominator in square
brackets. Therefore, 
\begin{equation}
\left\langle \left| x\right| ^{q}\right\rangle =C(q,\alpha _{1})\left( \frac{%
t}{\alpha _{1}}\right) ^{q/\alpha _{1}},\qquad q<\alpha _{1},  \label{5.23}
\end{equation}
where $C(q,\alpha _{1})$ is given by Eq.(\ref{3.13}), and 
\begin{equation}
\delta x\sim \left\langle \left| x\right| ^{q}\right\rangle ^{1/q}\propto
t^{1/\alpha _{1}}  \label{5.24}
\end{equation}
for $t\rightarrow 0$.

In order to get the $q$-th moment at large $t$, we again use the Laplace
method. Turn to Eq.(\ref{5.13}), where $\phi (k)$ is given by Eq.(\ref{5.21}%
) and 
\begin{equation}
\zeta (k)=\frac{\phi ^{\prime }(k)}{k^{q}}=\frac{1}{k^{1+q}}\frac{%
a_{1}\alpha _{1}k^{-\alpha _{1}}+a_{2}\alpha _{2}k^{-\alpha _{2}}}{\left(
a_{1}k^{-\alpha _{1}}+a_{2}k^{-\alpha _{2}}\right) ^{2}}.  \label{5.25}
\end{equation}
Since at small $k$ we have $\phi (k)\approx k^{\alpha _{2}}$, $\zeta
(k)\approx \left( \alpha _{2}/a_{2}\right) k^{\alpha _{2}-q-1}$, then at
large $t$ we again have exactly the case of Laplace asymptotic integral
\cite{Olv74}. For $J(t)$ we have
immediately $J(t)=a_{2}^{-q/\alpha _{1}}t^{q/\alpha _{2}-1}\Gamma
(1-q/\alpha _{2})$, and 
\begin{equation}
\left\langle \left| x\right| ^{q}\right\rangle \approx C(q,\alpha
_{2})\left( \frac{t}{\alpha _{2}}\right) ^{q/\alpha _{1}},\qquad q<\alpha
_{1},  \label{5.26}
\end{equation}
where $C(q,\alpha )$ is given by Eq.(\ref{3.13}), and 
\begin{equation}
\delta x\sim \left\langle \left| x\right| ^{q}\right\rangle ^{1/q}\propto
t^{1/\alpha _{2}}  \label{5.27}
\end{equation}
for $t\rightarrow \infty $.

Therefore, at small times the characteristic displacement grows as $%
t^{1/\alpha _{1}}$, whereas at large times it grows as $t^{1/\alpha _{2}}$.
Since $\alpha _{1}<\alpha _{2}$, we encounter retarding superdiffusion.

\section{Conclusions}

Distributed-order diffusion equations generalize the approach based on
fractional diffusion equations to processes lacking
temporal scaling. The typical forms of fractional diffusion equations can be
classified with respect to the position of the fractional derivative in
time/coordinate instead of or in addition to the first and second derivatives
in the classical Fick's form. In the present paper we considered the
corresponding forms of distributed-order fractional diffusion equations and
elucidate the effects described by different classes of such equations. We
show that equations with the distributed-order fractional operator
replacing the corresponding whole-number derivative describe processes
getting more anomalous in course of the time, i.e. the accelerating
superdiffusion or retarded subdiffusion. On the opposite, equations
with additional fractional operators on the ''wrong'' side of the
diffusion equation describe processes getting less anomalous, i.e.
retarded superdiffusion and accelerating subdiffusion.

\section{Acknowledgements}

The work was supported by an INTAS grant. IMS gratefully acknowledges the
support by the Fonds der Chemischen Industrie.

\end{document}